\documentclass[12pt]{article}
\usepackage{amsmath,amssymb}
\usepackage{bm}

\makeatletter
  
  \@addtoreset{equation}{section}
\makeatother

\begin{document}
\baselineskip 16pt 
\allowdisplaybreaks

\thispagestyle{empty}
\vspace*{4.5cm}
\begin{center}
\Large {\bf 
On broken zero modes of a string world sheet, \\
and a correlation function of a 
1/4 BPS Wilson loop and a 1/2 BPS local operator
\\}

\vspace{1.5cm}

\normalsize
 \vspace{0.4cm}

Akitsugu {\sc Miwa}\footnote{e-mail address:\ \ 
{\tt miwa.akitsugu@nihon-u.ac.jp}}

\vspace{0.7cm}

{\it 
Department of Physics, College of Science and Technology, 
Nihon University, \\
1-8-14, Kanda-Surugadai, Chiyoda-ku, Tokyo 101-8308, Japan} \\

\vspace{3cm}
{\bf Abstract} 
\end{center}
We reconsider a gravity dual of a 1/4 BPS Wilson loop.
In the case of an expectation value of the Wilson loop, 
it is known that broken zero modes of a string world sheet 
in the gravity side play important roles in the limit 
$\lambda \to \infty$\, with keeping the combination 
$\lambda \cos^2 \theta_0$ finite. 
Here, $\lambda$ is the 't Hooft coupling constant 
and $\theta_0$ is a parameter of the Wilson loop.
In this paper, we reconsider a gravity dual 
of a correlation function between the Wilson loop
and a 1/2 BPS local operator with R charge $J$\,.
We take account of contributions coming 
from the same configurations of the above-mentioned 
broken zero modes.
We find an agreement with the gauge theory side
in the limit $J \ll \sqrt{\lambda \cos^2 \theta_0} $\,.

\vspace{2em}

\newpage

\setcounter{page}{1}

\tableofcontents

\section{Introduction}

In the context of the AdS/CFT correspondence 
\cite{Maldacena:1997re}\cite{Gubser:1998bc}\cite{Witten:1998qj},
dual objects of Wilson loop operators are 
given in terms of string world sheets
attached to the loops on the AdS boundary 
\cite{Maldacena:1998im}\cite{Rey:1998ik}.
An expectation value of the Wilson loop in the gauge theory side, 
for example, is conjectured to be equal to a string disk partition function
which is, in principle, given by a path integral 
of the string world sheet.
In literature, the path integral is usually approximated 
by using a classical solution, and then a strong coupling limit 
of the gauge theory result is reproduced.
A natural way to go beyond the classical string would be 
taking account of effects coming from small fluctuations 
around the classical solutions perturbatively.
For developments in such a method, see the papers \cite{Forste:1999qn}\cite{Drukker:2000ep}\cite{Kruczenski:2008zk}\cite{Kristjansen:2012nz}.

Besides such analyses, 
an interesting method has been proposed in \cite{Drukker:2006ga}
for the case of an expectation value of a 1/4 BPS Wilson loop operator.
The operator depends on a parameter $\theta_0$\,, 
and for a special case with $\theta_0 = \pi / 2$\,,
there is a three-parameter family of string solutions 
corresponding to a single common Wilson loop. 
These three parameters correspond to zero modes
of the string world sheet.
An integration of the exact zero modes just
gives a multiplicative factor.
The author of the paper \cite{Drukker:2006ga} further considered
a limit in which $\theta_0$ is not exactly equal to $\pi/2$\,,
but very close to it.
In such a limit, the zero modes
get a potential which is not exactly flat.
Hence they are no longer exact zero modes.
However, these modes are expected to be still much lighter
than generic fluctuations in the limit 
$\lambda \to \infty$, $\theta_0 \to \pi/2$\,.
Then the integration over such modes gives non-negligible contributions.
In fact, it was found that by keeping the combination
$\lambda' \equiv \lambda \cos^2 \theta_0$ finite, 
the integration over such broken zero modes
reproduces the exact gauge theory result 
of the Wilson loop expectation value 
in the planar limit \cite{Drukker:2006ga}.

In this paper we apply the method of the broken zero modes 
to a gravity dual of a correlation function between
the 1/4 BPS Wilson loop operator and a 1/2 BPS 
local operator with an R charge $J$\,.
The same system has been studied in 
\cite{Semenoff:2006am}\cite{Giombi:2012ep}\cite{Enari:2012pq}
based on the method developed in earlier works 
\cite{Berenstein:1998ij}\cite{Zarembo:2002ph}
by treating the string world sheet classically.\footnote{
In \cite{Zarembo:2002ph}, a correlation function 
between a 1/2 BPS Wilson loop and the 1/2 BPS local operator 
is studied by taking account of small fluctuations
around a classical solution. 
It is found that such an analysis 
reproduces the gauge theory result in the limit
$\lambda \to \infty$ with keeping $J^2/\sqrt{\lambda}$ finite.
A relation between our result and the one in \cite{Zarembo:2002ph}
is discussed in section \ref{Correlation function}.
}
We start with reconsidering the broken zero modes 
by giving an explicit form of them, and we check that 
the integration over the modes reproduces 
the result of \cite{Drukker:2006ga}. 
Then we apply the method 
to the case of the correlation function.
We find that our computation reproduces 
the gauge theory result approximately only 
in the limit $J \ll \sqrt{\lambda'}$\,. 
Deviations in other range is actually 
natural because our string configuration 
does not carry any angular momentum
which corresponds to the R charge $J$ of the local operator. 
Hence only in the limit $J \ll \sqrt{\lambda'}$\,,
the configuration is acceptable.
We will see that the limit still allows us
to go beyond the purely classical limit.

The paper is organized as follows. 
In section 2, we introduce the 1/4 BPS Wilson loop operator 
and review some known facts about the operator and its gravity dual. 
In section 3, we discuss an explicit form of the broken zero modes
and show that their contributions reproduce the planar limit of the
Wilson loop expectation value as discussed in \cite{Drukker:2006ga}.
In section 4, we apply the method of the broken zero modes 
to the correlation function of the 1/4 BPS Wilson loop 
and the 1/2 BPS local operator. 
Section 5 is devoted to a summary and discussions. 

\section{A review of 1/4 BPS Wilson loop operators}

\subsection{1/4 BPS Wilson loop operators  and modified Bessel functions}
A Wilson loop operator in the ${\cal N}=4$
super Yang-Mills (SYM) theory is characterized 
by the shape of the loop $x_i ( \tau )$ ($i = 1 \ldots 4$)
in the Euclidean four-dimensional space and also 
another set of ``coordinates'' $\Theta_I( \tau )$ ($I=1\ldots 6$),
which specifies how the operator depends on the 6 scalar fields $\Phi_I$
\cite{Maldacena:1998im}\cite{Rey:1998ik}:
\begin{equation}
 W(C) = {1 \over N}
{\rm trP} \exp
\oint 
\Big(
i A_i(\vec x(\tau)) \dot x_i(\tau)
+
|\dot {\vec x}(\tau)| \Theta_I(\tau) \Phi_I(\vec x(\tau)) 
\Big) d \tau \,.
\label{W(C)}
\end{equation}
In this paper, we take the following choice \cite{Drukker:2006ga}:
\begin{align}
\vec x( \tau ) & = 
(
\, a \cos \tau \,,\,a \sin \tau \,,\, 0 \,, \,0 \,
)\,, 
\label{x} \\
\vec \Theta( \tau ) & =
( \,
\sin \theta_0 \cos \tau \,, \,
\sin \theta_0 \sin \tau \,, \,
\cos \theta_0 \,, \,
0 \,, \,
0 \,, \,
0 \,
)\,. 
\label{Theta}
\end{align}
The operator depends on two parameters $a$ and $\theta_0$.
We assume that the range of $\theta_0$ is $0 \leq \theta_0 \leq \pi/2 $\,.
The operator preserves 1/4 
of the supersymmetries in the SYM theory
for a generic value of $\theta_0$\,.

The planar limit of the expectation value 
of the operator is studied in \cite{Drukker:2006ga} based 
on the analysis of \cite{Erickson:2000af}\cite{Drukker:2000rr}.
The result is given in terms of 
the modified Bessel function $I_1$ as follows:
\begin{align}
\langle W(C) \rangle 
& = 
{2 \over \sqrt{\lambda '}} I_1 (\sqrt{\lambda'})\,.
\label{<W>}
\end{align}
Here, as mentioned in the introduction, 
$\lambda'$ is related to 
the 't Hooft coupling constant $\lambda$ as
$\lambda' = \lambda \cos^2 \theta_0$\,.

\subsection{String solutions in the gravity side}

In the gravity side, counterparts of \eqref{<W>} 
is formally given in terms of a string path integral as follows:
\begin{align}
\langle W(C) \rangle
& \quad \leftrightarrow \quad 
\int {\rm e}^{-S}\,. 
\label{<W>=e^S}
\end{align}
Here, the right hand side (the gravity side) is
expressed only symbolically.
$S$ is a total string world sheet action 
which includes a boundary term.
We assume that the bulk part of it is 
the Polyakov type in the conformal gauge:
\begin{equation}
S_{\rm bulk} = {\sqrt \lambda \over 4 \pi}
\int d \tau d \sigma g_{MN} \partial_a 
{\cal X}^M (\tau,\sigma) \partial_a {\cal X}^N (\tau, \sigma)\,.
\end{equation}
Here $M$ and $N$ are ten-dimensional indices and $a$ is a world sheet index.
$g_{MN}$ are the components of the metric for the AdS$_5 \times$S$^5$ background,
while ${\cal X}^M(\tau, \sigma)$ are string coordinates
which are functions of world sheet coordinates ($\tau$, $\sigma$)\,.
For simplicity, we use the unit in which the common radius of the 
AdS$_5$ and the S$^5$ is taken to be $1$, 
then the 't Hooft coupling constant $\lambda$ and the Regge slope 
$\alpha'$ are related by $\sqrt \lambda = 1/\alpha'$\,.
As for the boundary term, 
we assume the one proposed in \cite{Drukker:1999zq}
throughout this paper.
In order to avoid unimportant complexities,
detailed computations of the boundary terms 
are summarized in appendix \ref{boundary terms}. 
The symbol $\int$ in \eqref{<W>=e^S} expresses the path integral 
of the string world sheet, 
the precise definition of which is beyond the scope of the present paper.
The boundary conditions for the AdS$_5$ coordinates, 
other than the radial direction, are given by \eqref{x},
while those for the S$^5$ are given by \eqref{Theta}.

In \cite{Drukker:2006ga}, 
the correspondence \eqref{<W>=e^S} 
for a generic value of $\theta_0$ is studied 
by evaluating the path integral 
at solutions for equations of motion. 
Then the large $\lambda '$ limit of 
the modified Bessel function is correctly reproduced. 
Let us briefly review the string solutions.

\subsubsection{Classical solution: AdS$_5$-part}
Let $\vec X = (X_0, X_1, X_2, X_3, X_4, X_5)$ be coordinates 
of a flat {\bf R}$^{1,5}$ space with a line element 
\begin{equation}
ds^2 = d \vec X \cdot d \vec X = 
- dX_0^2 + dX_1^2 + dX_2^2 + dX_3^2 + dX_4^2 + dX_5^2\,.
\end{equation}
Here, the inner product ``\,$\cdot$\,'' 
for $\vec X$ is defined with the signatures $(-,+,+,+,+,+)$\,.
Then the Euclidean AdS$_5$ can be expressed as a hypersurface
defined by the following equation:
\begin{equation}
\vec X \cdot \vec X = 
-X_0^2 + X_1^2 + X_2^2 + X_3^2 + X_4^2 + X_5^2 = -1\,.
\end{equation}

An interesting feature of the string solutions dual to the
Wilson loop \eqref{W(C)} with \eqref{x} and \eqref{Theta}
is that the AdS$_5$-part of it is 
independent of the parameter $\theta_0$\,.
This means that it is the same configuration as the 1/2 BPS Wilson loop
\cite{Berenstein:1998ij}\cite{Drukker:1999zq},  
which corresponds to the case with $\sin \theta_0 = 0$.
The explicit form of the solution in this coordinate system
is given by
\begin{equation}
X_0 = \coth \sigma\,, \quad 
X_1 = {\rm cosech} \sigma \cos \tau\,, \quad 
X_2 = {\rm cosech} \sigma \sin \tau\,, \quad 
X_3  = X_4 = X_5 = 0\,.
\label{AdS-SOL}
\end{equation}
Here, ranges of the world sheet coordinates ($\tau, \sigma$)
are $0\leq \tau < 2 \pi$ and $0 \leq \sigma \leq \infty$\,, 
respectively.
The solution satisfies the following equations:
\begin{equation}
\partial_\tau \vec X \cdot \partial_\tau \vec X
=
\partial_\sigma \vec X \cdot \partial_\sigma \vec X 
=
{\rm cosech}^2 \sigma\,, \quad
\partial_\tau \vec X \cdot \partial_\sigma \vec X = 0\,.
\label{Virasoro-AdS}
\end{equation} 
Hence, the Virasoro constraints need to be 
``satisfied separately'' by the AdS$_5$-part and the S$^5$-part.

In order to study the AdS/CFT correspondence, 
it is convenient to introduce 
the Poincar\'e coordinate $(z,\vec x)=(z,x_1, x_2, x_3, x_4)$
by the following coordinate transformation:
\begin{align}
X_0  = {z^2 + {(\vec x)}^2 + a^2 \over 2az}\,, \quad
X_5 = {z^2 + {(\vec x)}^2 - a^2 \over 2az}\,, \quad 
X_i = { x_i \over z} \,\,\, (i=1\ldots 4)\,,
\end{align}
where ${(\vec x)}^2 = x_1^2 + x_2^2 + x_3^2 + x_4^2$\,.
The line element in this coordinate is given by
\begin{equation}
ds^2 = d \vec X \cdot d \vec  X
= {dz^2 + {(d \vec x)}^2 \over z^2}\,,
\end{equation}
and the solution \eqref{AdS-SOL} is expressed as follows:
\begin{align}
z = a \tanh \sigma \,, \quad 
x_1 = a \,{\rm sech}\, \sigma \cos \tau \,, \quad
x_2 = a \,{\rm sech}\, \sigma \sin \tau \,, \quad
x_3 = x_4 = 0\,.
\label{AdS-SOL-3}
\end{align}
It is attached to the path \eqref{x} on the AdS$_5$ boundary $z=0$\,,
at $\sigma=0$\,.

\subsubsection{Classical solutions: S$^5$-part}

Let $\vec Y = (Y_1,Y_2,Y_3,Y_4,Y_5,Y_6)$ be coordinates 
of a flat {\bf R}$^6$ space with a line element
\begin{equation}
ds^2 
= d\vec Y \cdot d \vec Y\,
= dY_1^2 + dY_2^2 + dY_3^2 + dY_4^2 + dY_5^2 + dY_6^2\,.
\end{equation}
The inner product ``\,$\cdot$\,'' for $\vec Y$ is 
defined with the all positive signatures.
The S$^5$ we consider is embedded in the 
space by the following equation:
\begin{equation}
\vec Y \cdot \vec Y =Y_1^2 + Y_2^2 + Y_3^2 + Y_4^2 + Y_5^2 + Y_6^2 = 1\,.
\end{equation}
For a generic value of $\theta_0$\,, there are two string solutions
corresponding to the Wilson loop:
\begin{equation}
\begin{split}
Y_1&  = {\rm sech}(\sigma_0 \pm \sigma) \cos \tau\,, \quad 
Y_2  = {\rm sech}(\sigma_0 \pm \sigma) \sin \tau\,, \quad 
Y_3  = \tanh(\sigma_0 \pm \sigma)\,, \\
Y_4&  = Y_5 = Y_6 = 0\,.
\end{split}
\label{no0S}
\end{equation}
Here, the parameter $\sigma_0$ is related to $\theta_0$ by
$\tanh \sigma_0 = \cos \theta_0$\,,
and the following boundary conditions 
are satisfied:
\begin{equation}
 \vec Y\big|_{\sigma = 0}
 =
 (\, 
\sin \theta_0 \cos \tau \,,\, 
\sin \theta_0 \sin \tau \,,\,
\cos \theta_0 \,, \,
0 \,,\,
0 \,,\,
0 
\,)\,. 
\label{bdy-cond}
\end{equation}
As mentioned below \eqref{Virasoro-AdS}, 
each configuration of \eqref{no0S} satisfies
the following S$^5$-part of the Virasoro constraints:
\begin{equation}
\partial_\tau \vec Y \cdot \partial_\tau \vec Y 
=
\partial_\sigma \vec Y \cdot \partial_\sigma \vec Y 
=
{\rm sech}^2(\sigma_0 \pm \sigma)\,, \quad
\partial_\tau \vec Y \cdot \partial_\sigma \vec Y = 0\,.
\end{equation}

\subsection{Zero modes and broken zero modes of the string world sheet}
One of interesting developments in \cite{Drukker:2006ga}
about the 1/4 BPS Wilson loop is 
that the exact integral representation of 
the modified Bessel function is reproduced 
by taking account of broken zero modes.
In the rest of this section, 
we focus on the S$^5$-part
and discuss the idea of the broken zero modes.
As for the AdS$_5$-part of the configuration,
we assume \eqref{AdS-SOL-3}.

\subsubsection{Zero modes in the case $\boldsymbol{\theta_0 = \pi/2}$}
Let us consider the case with $\cos \theta_0 = 0$\,, 
i.e., $\theta_0 = \pi / 2$\,.
In this case, the boundary conditions for the S$^5$ coordinates 
are given by the following great circle on the $Y_1Y_2$-plane:
\begin{equation}
\vec Y\big|_{\sigma=0} =  
( \, \cos \tau \,, \, \sin \tau \, , \, 0 \, , \, 0 \, , \, 0 \, ,
\,0 \,)\,.
\end{equation}
There is a three-parameter family of string solutions
satisfying the boundary condition.
It is given by the following configuration 
\begin{align}
\begin{split}
 Y_1 & = {\rm sech}\, \sigma \cos \tau\,, \\
 Y_2 & = {\rm sech}\, \sigma \sin \tau\,, \\
 Y_3 & = \tanh \sigma \cos \alpha\,, \\
 Y_4 & = \tanh \sigma \sin \alpha \cos \beta\,, \\
 Y_5 & = \tanh \sigma \sin \alpha \sin \beta 
\cos \gamma\,, \\
 Y_6 & = \tanh \sigma \sin \alpha \sin \beta
\sin \gamma\,. 
\end{split}
\label{Zero modes}
\end{align}
Here, $\alpha$, $\beta$ and $\gamma$ are the three constant parameters, 
the ranges of which are $0 \leq \alpha \leq \pi$\,, 
$0 \leq \beta \leq \pi$
and $0 \leq \gamma < 2 \pi$\,.
The solution \eqref{no0S}, with $\sigma_0 = 0$\,,
corresponds to the configuration \eqref{Zero modes}
with $\alpha=0\,, \, \pi$\,.
These three parameters correspond to zero modes, i.e., the string
action does not depend on them.
In fact, the total action including the boundary term 
for these configurations turns out to be zero.
Then, if we approximate the string path integral 
by integrating over only these zero modes, 
we obtain 
\begin{equation}
\int {d \Omega_3 \over 2 \pi^2}
{\rm e}^{- S |_{\text{0-modes}} }
=1,
\qquad \big(\, S|_{\text{0-modes}} = 0 \,\big).
\end{equation}
Here, the integral measure is defined by 
$d \Omega_3 =  d\alpha d \beta d \gamma \sin^2 \alpha \sin \beta$
and we have fixed the normalization so that the gauge theory result, 
$\langle W(C) \rangle = 1$\,, is reproduced.
We use this unique normalization throughout this paper.
This special case of the Wilson loop and its gravity dual 
are first studied in \cite{Zarembo:2002an}.
The existence of the three zero modes and also 
the fact that string total action becomes zero 
are found in that paper.

\subsubsection{Broken zero modes for  
$\boldsymbol{\theta_0 \sim \pi/2}$}

For a generic value of $\theta_0$\, 
there are only two solutions 
which are given by \eqref{no0S},
and there are no zero modes around the solutions.
This is because the non-vanishing third component 
$\cos \theta_0$ in \eqref{bdy-cond} breaks the S$^3$ symmetry.
Although the boundary condition has S$^2$ symmetry, 
rotations in $Y_4Y_5Y_6$-space do not generate 
any independent solutions since 
the world sheet \eqref{no0S}
is sitting at the origin of that space.
Evaluating the string path integral 
by the classical solution \eqref{no0S}, 
the saddle point values of the modified Bessel function, 
${\rm e}^{\pm \sqrt{ \lambda'}}$\,, are reproduced \cite{Drukker:2006ga}.

Beyond such classical analyses, 
a remarkable success in the limit $\theta_0 \sim \pi/2$ 
is given in \cite{Drukker:2006ga}.
There, it is argued that for a small deviation 
from the S$^3$ symmetric case of $\theta_0 = \pi/2$\,,
the symmetry is broken only slightly,
and the modes corresponding to the zero modes 
in the symmetric case should still have significant 
contributions to the string path integral.
By following \cite{Drukker:2006ga}, 
we call such modes as the ``broken zero modes''.
In fact, it is reported if we take the limit 
$\lambda \to \infty $ and $\cos \theta_0 \to 0$ 
with keeping the combination $\lambda' = \lambda \cos^2 \theta_0$
finite, then the broken zero modes give rise to a finite potential 
$ S|_{\text{/\!\!\!0-modes}}  = - \cos \alpha \sqrt{\lambda'}$\,.\footnote{
Note that our parameter $\alpha$ 
is related to the one in the paper \cite{Drukker:2006ga}
by $\alpha \leftrightarrow \pi- \alpha$.
}
By taking account of only these broken zero modes, 
the string path integral is approximated as
\begin{equation}
\int {d \Omega_3 \over 2 \pi^2}
{\rm e}^{- S|_{\text{/\!\!\!0-modes}} }
=
{2 \over \pi} 
\int d \alpha \sin^2 \alpha {\rm e}^{\cos \alpha \sqrt{\lambda'}}
=
{2 \over \sqrt{\lambda'}} I_1 (\sqrt{\lambda'})\,. 
\label{Bessel}
\end{equation}
This agrees with the planar limit 
of the Wilson loop expectation value.

An important point here is that it is {\it not} 
a finite $\lambda$ result, 
but the large $\lambda$ limit is taken.
This implies that we do not need to care about 
corrections coming from other generic string fluctuations,
since they are expected to be suppressed by 
inverse powers of $ \sqrt \lambda$\,.
Note also that small corrections which are of higher orders 
with respect to $\cos \theta_0$ are suppressed since
$\cos \theta_0 = \sqrt{\lambda'} / \sqrt{\lambda}$\, 
and $\lambda'$ is now finite.

\section{More about the broken zero modes}
In this section, we discuss an explicit form of the
S$^5$-part of the broken zero modes and reproduce 
the result of \cite{Drukker:2006ga}.
We also give a more systematic argument 
by assuming a symmetric ansatz. 
As for the AdS$_5$-part we assume the same configuration \eqref{AdS-SOL-3}.

\subsection{An explicit form of the broken zero modes}
The idea is to construct a three-parameter family 
of string configurations satisfying the common 
boundary condition \eqref{bdy-cond}.
In the case $\theta_0 = \pi /2$\, 
the three parameters should be identical to 
those in \eqref{Zero modes}.
So, let us call them $\alpha$, $\beta$ and $\gamma$\,.
We also expect that the configuration reduces 
to the solution \eqref{no0S} for specific values 
of the parameter, i.e., $\alpha=0\,,\,\pi$\,.

It is not difficult to check that the following
configuration satisfies all of these requirements:
\begin{align}
\begin{split}
 Y_1 & = Y_1 (\tau,\sigma,\alpha)= f(\sigma,\alpha) \cos \tau\,, \\
 Y_2 & = Y_2 (\tau, \sigma,\alpha) = f(\sigma,\alpha) \sin \tau\,, \\
 Y_3 & = Y_3 (\sigma, \alpha) =
f(\sigma,\alpha) (\cosh \sigma_0 \sinh\sigma \cos \alpha + \sinh \sigma_0 \cosh \sigma)\,,\\
 Y_4 & = Y_4(\sigma, \alpha, \beta) =
f(\sigma,\alpha) \sinh \sigma \sin \alpha \cos \beta\,, \\
 Y_5 & = Y_5(\sigma, \alpha, \beta, \gamma) =
f(\sigma,\alpha) \sinh \sigma \sin \alpha \sin \beta \cos \gamma\,, \\
 Y_6 & = Y_6(\sigma, \alpha, \beta, \gamma) =
f(\sigma,\alpha) \sinh \sigma \sin \alpha \sin \beta \sin \gamma\,,
\end{split}
\label{BZ}
\end{align}
where $f(\sigma,\alpha)$ is defined by 
\begin{equation}
 f(\sigma,\alpha) 
= 
{1 \over \cosh \sigma_0 \cosh \sigma + \sinh \sigma_0 \sinh \sigma \cos
\alpha}\,.
\label{f}
\end{equation}
The configuration also satisfies the following equations
\begin{equation}
 \partial_\tau \vec Y \cdot \partial_\tau \vec Y
 = 
 \partial_\sigma \vec Y \cdot \partial_\sigma \vec Y
 = 
 f^2 (\sigma,\alpha)\,, \quad
 \partial_\tau \vec Y \cdot \partial_\sigma \vec Y = 0\,.
\label{S-Virasoro}
\end{equation}
These equations mean that the Virasoro constraints are
satisfied by the S$^5$-part separately
from the AdS$_5$-part.
Hence, the configuration \eqref{AdS-SOL-3} 
for the AdS$_5$-part can be used without any change.

\subsection{$\bm \alpha$\,, $\bm \beta$\,, $\bm \gamma$  as S$^5$ coordinates}
Although it is not clear, \eqref{BZ} with \eqref{f}
defines a coordinate system $(\tau, \sigma, \alpha, \beta, \gamma)$ 
of an S$^5$, which covers the entire S$^5$ if we take the ranges as follows:
 \begin{equation}
 0 \leq \tau < 2 \pi\,, \quad
 0 \leq \sigma \leq \infty\,, \quad
 0 \leq \alpha \leq \pi\,, \quad
 0 \leq \beta \leq \pi\,, \quad
 0 \leq \gamma < 2 \pi\,.
 \end{equation}
This property assures that different sets of 
the parameters $( \alpha\,,\,\beta\,,\,\gamma)$
give different string configurations, i.e., 
they are not related through any redefinition of
world sheet coordinates $(\tau\,, \,\sigma)$\,.

Let us confirm the property briefly.
First we introduce the radial coordinate $R$
in the $Y_4Y_5Y_6$-space:
\begin{equation}
R=\sqrt{Y_4^2 + Y_5^2 + Y_6^2}\,.
\end{equation}
For \eqref{BZ}, 
it is given by 
\begin{equation}
R(\sigma, \alpha) 
= 
f(\sigma,\alpha)\sinh \sigma \sin \alpha\,,
\end{equation}
and we find the following relation:
\begin{equation}
 \tan \alpha Y_3(\sigma,\alpha) 
 - 
 {\rm sech}\,\sigma_0 R(\sigma,\alpha) 
 = \tanh \sigma_0 \tan \alpha\,,
 \label{Y3-R}
\end{equation}
where the function $Y_3(\sigma,\alpha)$ is defined in \eqref{BZ}.
For a given value of $\alpha$\,,
$R(\sigma,\alpha)$ is a monotonically increasing function of $\sigma$\,.
Hence, as $\sigma$ increases from $0$ to $\infty$\,, 
the point $( Y_3\,, R ) = ( Y_3(\sigma,\alpha)\,, R(\sigma,\alpha) )$
moves monotonically along the line segment between the following two points:
\begin{equation}
 \sigma = 0 :
 \begin{cases}
  Y_3  = \tanh \sigma_0\,, \\
  R = 0\,,
 \end{cases} 
 \; \sigma = \infty :
 \begin{cases}
  Y_3  = \overline Y_3 (\alpha) \equiv
 \displaystyle
  {
   \cosh \sigma_0 \cos \alpha + \sinh \sigma_0 
  \over 
  \cosh \sigma_0 + \sinh \sigma_0 \cos \alpha}\,, \\[2mm]
 R = \overline R (\alpha) \equiv
 \displaystyle{ \sin \alpha 
 \over 
 \cosh \sigma_0 + \sinh \sigma_0 \cos \alpha}\,.
 \end{cases}
\end{equation}
On the other hand, if we increase $\alpha$ 
from $0$ to $\pi$\,, 
the end point $(Y_3,R)=(\overline Y_3 (\alpha)\,, \overline R(\alpha))$ 
of the line segment moves, again monotonically, 
on the semi-circle which is defined by 
$ Y_3^2 +  R^2 = 1$ and $ R \geq 0$\,.
More precisely, $\overline Y_3(\alpha)$ 
is a monotonically decreasing function of $\alpha$\,.
This means that by changing $\sigma$ and $\alpha$\,,
the point $(Y_3\,, R) = (Y_3(\sigma\,, \alpha)\,, R(\sigma\,, \alpha))$ 
covers the half disk
$Y_3^2 + R^2 \leq 1$, $R \geq 0$ exactly once.
Next, if we choose some specific point on the half disk, 
then the corresponding slice of the S$^5$ is given by a 
direct product of an S$^1$ on the $Y_1Y_2$-plane
and an S$^2$ in the $Y_4Y_5Y_6$-space, 
which are parametrized by the independent parameters $\tau$ 
and ($\beta$\,, $\gamma$)\,, respectively.
Hence the entire S$^5$ is covered.

If we consider ($\tau, \sigma, \alpha, \beta, \gamma$)
as the S$^5$ coordinate system, the line element is given as follows:
\begin{equation}
 ds^2  = d \vec Y \cdot d \vec Y
 = f^2(\sigma, \alpha)
\Big[
d \tau^2 + d \sigma^2
+
\sinh^2 \sigma 
\big( 
d\alpha^2 + \sin^2 \alpha (d \beta^2 + \sin^2 \beta d \gamma^2)
\big)
\Big]\,.
\end{equation}
From this expression, we understand that
if we consider ($\tau$, $\sigma$) as the world sheet coordinates 
and ($\alpha$, $\beta$, $\gamma$) as the constant parameters, 
then \eqref{S-Virasoro} is satisfied.

\subsection{The broken zero modes and the modified Bessel function}

By using \eqref{AdS-SOL-3} and \eqref{BZ},
the total string action for the broken zero modes,
$S|_{\text{/\!\!\!0-modes}}$,  
including the boundary term $S_{\rm boundary}$ 
is evaluated as follows:
\begin{align}
 S|_{\text{/\!\!\!0-modes}}
& = {\sqrt \lambda \over 4 \pi} 
\int_0^{2\pi} 
d \tau 
\int_{\sigma_{\rm min}}^\infty 
d \sigma 
\big( 
\partial_\tau \vec X \cdot \partial_\tau \vec X 
+
\partial_\sigma \vec X \cdot \partial_\sigma \vec X 
+
\partial_\tau \vec Y \cdot \partial_\tau \vec Y 
+
\partial_\sigma \vec Y \cdot \partial_\sigma \vec Y 
\big) 
+ S_{\rm boundary} \notag \\
& = \sqrt \lambda \int_{\sigma_{\rm min}}^\infty d \sigma
\big( {\rm cosech}^2 \, \sigma + f^2(\sigma,\alpha) \big)
+ S_{\rm boundary} \notag \\
& =
\sqrt \lambda
\bigg[
- \coth \sigma 
+ {1 \over \cos \alpha}
{\tanh \sigma \cos \alpha + \tanh \sigma_0
\over
1 + \tanh \sigma_0 \tanh \sigma \cos \alpha}
\bigg]_{\sigma_{\rm min}}^\infty 
+ S_{\rm boundary} \notag \\
&
=
\sqrt \lambda \coth \sigma_{\rm min} 
-
\sqrt{\lambda} \tanh \sigma_0
\bigg[
\cos \alpha 
+ \tanh \sigma_0
{\sin^2 \alpha \over 1 + \tanh \sigma_0 \cos \alpha}
\bigg]
+ 
S_{\rm boundary}\,.
\label{S(a,b,c)-0}
\end{align}
In the final expression, the cutoff parameter $\sigma_{\rm min}$ 
is set to be $0$ except for the first term 
$\sqrt \lambda \coth \sigma_{\rm min}$\, 
(and also for the boundary term).
As explained in appendix \ref{boundary terms},
this divergent term is cancelled by the boundary term $S_{\rm boundary}$\,.
We take the limit $\sigma_0 \to 0$ and $\lambda \to \infty $ 
with keeping 
$\sqrt{\lambda'}  = \sqrt \lambda \tanh \sigma_0 
= \sqrt \lambda \cos \theta_0$ 
finite, and obtain
\begin{equation}
 S|_{\text{/\!\!\!0-modes}}
 =
 - \cos \alpha \sqrt{\lambda'} + {\cal O}(\tanh \sigma_0)
 =
 - \cos \alpha \sqrt{\lambda'} + {\cal O}(1/ \sqrt \lambda)\,.
\label{S(a,b,c)-1}
\end{equation}
This is the result given in \cite{Drukker:2006ga}
and the  modified Bessel function is reproduced as \eqref{Bessel}.

\subsection{Some general arguments for small $\sigma_0$}
In the previous subsection, we discussed the 
broken-zero-mode configurations in an ad hoc manner.
If we assume small $\sigma_0$ from the beginning,
it is also possible to restrict the configuration
more systematically.

Let us start with the following four assumptions
for the broken zero modes:
\begin{enumerate}
 \item The configurations reduce to the exact zero modes 
 \eqref{Zero modes} in the limit 
$\sigma_0 \to 0$.
 \item The configurations preserve S$^1$ symmetry 
 on the $Y_1Y_2$-plane corresponding to $\tau$.
 \item $(\beta, \gamma)$ describe flat directions 
 corresponding to the S$^2$ on the $Y_4Y_5Y_6$-space.
 \item The Virasoro constraints are satisfied 
 separately by the S$^5$-part and by the AdS$^5$-part.
\end{enumerate}
The first assumption is a necessary condition for the 
broken zero modes, while the others are just for simplification.
A more generic argument is beyond the scope of this paper.
The configurations which satisfy these assumptions
are written as follows:
\begin{align}
\begin{split}
 Y_1 & = 
 \Big(\,{\rm sech}\,\sigma + \tanh \sigma_0 y_1(\sigma,\alpha)
 + {\cal O}(\tanh^2 \sigma_0)
 \Big) \cos \tau\,, \\
 Y_2 & = 
 \Big(\,{\rm sech}\,\sigma + \tanh \sigma_0 y_1(\sigma,\alpha)
+ {\cal O}(\tanh^2 \sigma_0)
 \Big) \sin \tau\,, \\
 Y_3 & = \tanh \sigma \cos \alpha + \tanh \sigma_0 y_2(\sigma, \alpha)
 + {\cal O}(\tanh^2 \sigma_0) \,, \\
 Y_4 & = 
 \Big(\tanh \sigma \sin \alpha + \tanh \sigma_0 y_3(\sigma, \alpha)
 + {\cal O}(\tanh^2 \sigma_0) \Big)
 \cos \beta\,, \\
 Y_5 & = 
 \Big(\tanh \sigma \sin \alpha + \tanh \sigma_0 y_3(\sigma, \alpha)
 + {\cal O}(\tanh^2 \sigma_0)
 \Big)
 \sin \beta \cos \gamma\,, \\
 Y_6 & = 
 \Big(\tanh \sigma \sin \alpha + \tanh \sigma_0 y_3(\sigma, \alpha)
 + {\cal O}(\tanh^2 \sigma_0)
 \Big)
 \sin \beta \sin \gamma\,. 
\end{split}
\label{BZ-generic}
\end{align}
Here, $y_1(\sigma\,, \alpha)$\,, $y_2(\sigma\,,\alpha)$\,, 
$y_3(\sigma\,, \alpha)$ are undetermined functions.
The S$^5$ condition $\vec Y \cdot \vec Y = 1$ and
the Virasoro constraint for the S$^5$-part,
$\partial_\tau \vec Y \cdot \partial_\tau \vec Y 
= \partial_\sigma \vec Y \cdot \partial_\sigma \vec Y
$\,,
require\footnote{
Another constraint
$
\partial_\tau \vec Y \cdot \partial_\sigma \vec Y = 0
$
is satisfied.
}
\begin{align}
0 & 
= 
{\rm sech}\,\sigma y_1 
+ \tanh \sigma \cos \alpha y_2 
+ \tanh \sigma \sin \alpha y_3\,,  \\
0 & = 
{1 \over \cosh \sigma} y_1
+
{\sinh \sigma \over \cosh^2 \sigma} \partial_\sigma y_1
-
{1 \over \cosh^2 \sigma} \cos \alpha \partial_\sigma y_2
-
{1 \over \cosh^2 \sigma} \sin \alpha \partial_\sigma y_3\,.
\end{align}
The boundary condition \eqref{bdy-cond} requires
\begin{equation}
 y_1(0,\alpha) = 0\,, 
 \quad
 y_2(0,\alpha) = 1\,,
 \quad 
 y_3(0,\alpha) = 0\,.
\end{equation}
The solution for these conditions is  given by 
\begin{equation}
y_1(\sigma, \alpha) 
= 
- \cos \alpha {\sinh \sigma \over \cosh^2 \sigma}\,, \quad
\cos \alpha y_2(\sigma, \alpha)  
+
\sin \alpha y_3(\sigma, \alpha)
=
{\cos \alpha \over \cosh^2 \sigma}\,.
\label{y_1y_2}
\end{equation}
Although there remains one arbitrary function, 
which we can take $y_2(\sigma,\alpha)$ or $y_3(\sigma,\alpha)$
for example,
it does not affect the computation of the 
string action.\footnote{
Note also that this arbitrary function does not 
introduce any extra zero mode.
This is because it does not correspond to any finite deformation 
of the configuration in the small $\tanh \sigma_0$ limit.
}
In fact, by using \eqref{BZ-generic}  and \eqref{y_1y_2},
the S$^5$-part of the bulk action is computed 
as follows:
\begin{align}
 S_{\textrm{bulk,\,S$^5$}}|_{\text{/\!\!\!0-modes}}
 & =
 {\sqrt \lambda \over 4 \pi}
 \int_0^{2\pi} d \tau 
 \int_0^\infty d \sigma
 ( 
 \partial_\tau \vec Y \cdot \partial_\tau \vec Y 
 +
 \partial_\sigma \vec Y \cdot \partial_\sigma \vec Y
 ) \\
 & = 
 \sqrt \lambda 
 \int_0^\infty d \sigma 
 \bigg(
 {1 \over \cosh^2 \sigma} 
 +
 2\tanh \sigma_0
 {y_1(\sigma,\alpha) \over \cosh \sigma} 
 +
 {\cal O}(\tanh^2 \sigma_0)
 \bigg) \\
 & = 
 \sqrt \lambda
 \bigg[
 1 - \cos \alpha \tanh \sigma_0 + {\cal O}(\tanh^2 \sigma_0)
 \bigg]\,.
\end{align}
It correctly reproduces the S$^5$-part of \eqref{S(a,b,c)-0}.

\section{Correlation function between a Wilson loop and a local operator}
\label{Correlation function}
Next let us consider the effect of the broken zero modes 
on the correlation function of the 1/4 BPS Wilson loop \eqref{W(C)}
and the following local operator:
\begin{equation}
 {\cal O}_J = {(2\pi)^J \over \sqrt{J \lambda^J}}{\rm tr} (\Phi_3 + i \Phi_4)^J\,.
\end{equation}
The overall factor is taken
so that the large $N$ limit of the
two point function is normalized as 
\begin{equation}
\langle {\cal O}^\dag_J(\vec x_1) {\cal O}_J(\vec x_2) \rangle 
= {1 \over | \vec x_1 - \vec x_2 |^{2J} }\,.
\label{<O_JO_J>}
\end{equation}

The large $N$ limit of the correlation function is well studied
in the gauge theory side 
\cite{Semenoff:2001xp}\cite{Semenoff:2006am}\cite{Giombi:2012ep}. 
It is given in terms of the modified Bessel functions as follows
\cite{Semenoff:2006am}:
\begin{equation}
{\langle W(C) {\cal O}_J(\vec x) \rangle \over \langle W(C) \rangle }
=
{1 \over 2 N}
{a^J \over \ell^{2J}}\sqrt{J \lambda'} 
{I_J (\sqrt{\lambda'}) \over I_1 (\sqrt{\lambda'})}\,.
\label{<WO>/<W>}
\end{equation}
Here we assume that the position of the 
local operator is $\vec x = (0,0,0,\ell)$\,,
and also that the distance $\ell$ between the center 
of the loop $C$ and the position of the 
local operator is much larger than 
the radius $a$ of the loop, i.e., $a \ll \ell$.
It was found in \cite{Semenoff:2006am} that the system 
with the 1/4 BPS Wilson loop and the 1/2 BPS local operator
preserves 1/8 of the supersymmetries of the SYM theory.
We review the supersymmetries in appendix \ref{Supersymmetry}.

The gravity dual of the correlation function 
has been also well studied
\cite{Semenoff:2006am}\cite{Giombi:2012ep}\cite{Enari:2012pq}\cite{Berenstein:1998ij}\cite{Zarembo:2002ph}.
The basic idea is to consider 
a propagation of a bulk mode 
between the local operator and the Wilson loop,
both of which are assumed to be on the AdS boundary.
In order to explain the idea, 
let us review the result of \cite{Semenoff:2006am} 
in subsection \ref{Review of classical},\footnote{
In \cite{Semenoff:2006am}, a two point function of the 
Wilson loops is considered in order to compute the correlation function.
This is another standard approach \cite{Berenstein:1998ij}.
See \cite{Giombi:2006de}, for example, 
for the same approach as the present paper.
}
where the string world sheet 
dual to the Wilson loop is treated as classical.
Then in subsection \ref{Summing up broken zero modes}, 
we take account of the contribution from the broken zero modes.

\subsection{Review of an analysis without broken zero modes}
\label{Review of classical}
In literature, the correlation function in the gravity side is 
computed based on the following correspondence:
\begin{equation}
{
\langle W(C) {\cal O}_J(\vec x_J) \rangle
\over 
\langle W(C) \rangle
}
\quad \leftrightarrow \quad 
{
\sum
{\delta \over \delta s_0^J(\vec x_J)} 
{\rm e}^{- S[s_0^J(\vec x)] }
\big|_{s_0^J = 0,\, \text{classical}}
\over 
\sum {\rm e}^{-S[s_0^J(\vec x)] }
\big|_{s_0^J = 0,\, \text{classical}}
}\,.
\label{<WO>=d/dse^S}
\end{equation}
Here, the total string action $S[s_0^J(\vec x)]$ includes 
the same boundary term $S_{\rm boundary}$ as before
(see appendix \ref{boundary terms}),
and the bulk part of it is given by 
\begin{equation}
 S[s_0^J(\vec x)]_{\rm bulk} = {\sqrt \lambda \over 4 \pi}
\int d \tau d \sigma 
G_{MN} \partial_a {\cal X}^M \partial_a {\cal X}^N\,.
\end{equation}
As explained shortly, the geometry $G_{MN}$ 
depends on a source $s_0^J$ of the local operator ${\cal O}_J$.
Both the denominator and the numerator of \eqref{<WO>=d/dse^S}
are evaluated 
at $s_0^J =0$ and also at classical world sheets as indicated by 
``classical''.
The symbols $\sum$ in \eqref{<WO>=d/dse^S}
express summations over classical world sheet solutions.
In the present case, we have two solutions which are given in \eqref{no0S}.

The source $s_0^J (\vec x)$ is included 
in a small fluctuation $h_{MN}$ of the geometry 
$G_{MN}$ around the AdS$_5 \times$S$^5$ background $g_{MN}$
\begin{equation}
 G_{MN} = g_{MN} + h_{MN}\,.
\end{equation}
Equations of motion and constraints for such fluctuations
are diagonalized in \cite{Lee:1998bxa}, 
and the metric fluctuation corresponding to the local operator 
${\cal O}_J$ is given by 
\begin{align}
 h_{\mu \nu}^{\rm AdS} & = 
 \Big[
 - {6 J \over 5} g_{\mu \nu}^{\rm AdS} + {4 \over J+1} D_{(\mu} D_{\nu)} 
 \Big]
 s^J(\vec x,z) {\cal Y}_J\,, \\
 h_{\alpha \beta}^{\rm S} & =
 2 J g_{\alpha \beta}^{\rm S} s^J(\vec x,z) {\cal Y}_J\,.
\end{align}
Here, the ten-dimensional indices $M$\,, $N$ are decomposed into 
two sets; $\mu$\,, $\nu$ and $\alpha$\,, $\beta$\,.
The former set is the indices for the AdS$_5$-part 
and the latter is for the S$^5$-part. 
$g^{\rm AdS}_{\mu \nu}$ and $g^{\rm S}_{\alpha \beta}$ 
are the indicated components of the background metric $g_{MN}$\,,
while $h^{\rm AdS}_{\mu \nu}$ and $h^{\rm S}_{\alpha \beta}$ are their fluctuations.
The round bracket in $D_{(\mu} D_{\nu)}$ expresses
the symmetric traceless part, and 
${\cal Y}_J$ is the spherical harmonics
on the S$^5$, which is defined by
\begin{equation}
{\cal Y}_J = {1 \over 2^{J/2}} ( Y_3 + i Y_4 )^J\,.
\end{equation}
Here, $Y_3$ and $Y_4$ are the coordinates on the unit S$^5$\,.
The classical solution for the bulk field $s^J(\vec x\,, z)$ 
is written in terms of a boundary function $s_0^J(\vec x)$
by using the Green's function as
\begin{align}
  s^J(\vec x\,,z)
 & =
 \int d^4 x'
G(\vec x\,,z\,; \vec x') 
 s_0^J(\vec x')\,, \\
  G(\vec x, z; \vec x')
  & = 
 c \bigg(
 {z \over z^2 + (\vec x - \vec x')^2}
 \bigg)^J\,.
\end{align}
The constant $c$ is given by
\begin{equation}
 c = {2^{{J\over 2}-2} (J+1) \over N \sqrt J}\,, 
\end{equation}
which is chosen so that the two point function 
of the local operators, computed in the gravity side,
satisfies the same normalization 
as the gauge theory side \eqref{<O_JO_J>}.

In \eqref{<WO>=d/dse^S}, the derivative 
with respect to the source $s_0^J(\vec x_J)$ acts on $h_{MN}$ as\footnote{
As it is shown in appendix \ref{boundary terms}, 
the derivative of the boundary term vanishes.
}
\begin{equation}
{\delta \over \delta s_0^J(\vec x_J)}
{\rm e}^{-S[s_0^J(\vec x)] }
\bigg|_{s_0^J=0}
=
{\rm e}^{- S[0] }
\bigg[
- {\sqrt \lambda \over 4 \pi} 
\int d \tau d \sigma
\bigg(
{\delta \over \delta s_0^J(\vec x_J)}
h_{MN} 
\bigg)
\partial_a {\cal X}^M \partial_a {\cal X}^N
\bigg]
\,.
\label{dh/ds}
\end{equation}
The AdS$_5$-part and the S$^5$-part of the derivatives
are given as follows:
\begin{align}
 {\delta \over \delta s_0^J(\vec x_J)} h_{\mu \nu}^{\rm AdS} 
 & =
 \Big[
 - 
 {6J \over 5} g^{\rm AdS}_{\mu \nu}
 +
 {4 \over J+1} D_{(\mu}D_{\nu)}
 \Big] 
G(\vec x, z;\vec x_J)
 {\cal Y}_J  \notag \\
 & \sim 
2J \Big[ - g^{\rm AdS}_{\mu \nu} 
+ {2 \over z^2} \delta_\mu^z \delta_\nu^z \Big] 
c{z^J \over \ell^{2J}}{\cal Y}_J\,, \label{dh/ds-2}\\
 {\delta \over \delta s_0^J(\vec x_J)} h_{\alpha \beta}^{\rm S} 
 & = 2 J g^{\rm S}_{\alpha \beta}  G(\vec x, z; \vec x_J) 
 {\cal Y}_J
   \sim 2 J g^{\rm S}_{\alpha \beta} c {z^J \over \ell^{2J}} {\cal Y}_J\,.
 \label{dh/ds-3}
\end{align}
Here, we have assumed that the distance $\ell$ 
between the center of the loop and the position 
of the local operator is much larger than the
radius $a$ of the loop.

By using the solutions
\eqref{AdS-SOL-3} and \eqref{no0S}, 
the classical value of \eqref{dh/ds} is evaluated as 
\begin{align}
 & 
 {\rm e}^{\pm \sqrt{\lambda'}}
 \bigg[
 {\sqrt \lambda \over 2}
 c {4J \over \ell^{2J}}a^J
 \int_0^\infty d \sigma
 \bigg(
 {1 \over \cosh^2 \sigma}
 - 
 {1 \over \cosh^2 (\sigma_0 \pm \sigma)} 
 \bigg) \tanh^J \sigma  {\cal Y}_J
 \bigg]\,. \label{SY-equation}
\end{align}
Here, the spherical harmonics ${\cal Y}_J$ is now given by
\begin{equation}
{\cal Y}_J 
={1 \over 2^{J/2}} \tanh^J(\sigma_0 \pm \sigma)\,.
\end{equation}
For simplicity, we consider only the small $\sigma_0$ limit.
Then \eqref{SY-equation} is evaluated as follows:
\begin{align}
(\pm 1)^{J+1} {\rm e}^{\pm \sqrt{\lambda'}}
\sqrt{\lambda'}
4J c {a^J \over \ell^{2J}} {1 \over 2^{J/2}}
\int_0^\infty d\sigma {\tanh^{2J+1} \sigma \over \cosh^2 \sigma} 
=
(\pm 1)^{J+1}
{1 \over 2N}
{a^J \over \ell^{2J}}
\sqrt{ J \lambda'}
{\rm e}^{\pm \sqrt{\lambda'}}\,.
\label{S-Y}
\end{align}
This gives two contributions coming from the two string solutions. 
These are summed up in the numerator of \eqref{<WO>=d/dse^S}.
The denominator is the summation of ${\rm e}^{\pm \sqrt{\lambda'}}$\,.
Then \eqref{S-Y} correctly reproduces the prefactor 
of \eqref{<WO>/<W>} in the large $\lambda'$ limit \cite{Semenoff:2006am}.

\subsection{Summing up broken zero modes}
\label{Summing up broken zero modes}
Let us now try to take account 
of quantum corrections coming from the 
broken zero modes.
For this purpose, we further generalize the 
expression \eqref{<WO>=d/dse^S} so that the 
integration over these modes are included
\begin{equation}
{
\langle W(C) {\cal O}_J(\vec x_J) \rangle 
\over 
\langle W(C) \rangle
}
\quad \leftrightarrow \quad 
{
\int {d\Omega_3 \over 2\pi^2} 
{\delta \over \delta s_0^J(\vec x_J) }
{\rm e}^{-S[s_0^J(\vec x)]}
\big|_{{s_0^J = 0},\,\text{/\!\!\!0-modes}}
\over 
\int {d\Omega_3 \over 2\pi^2} 
{\rm e}^{-S[\alpha\,, \,\beta\,, \,\gamma\,;\, s_0^J(\vec x)]}
\big|_{s_0^J = 0,\,\text{/\!\!\!0-modes}}
}\,.
\label{<WO>=d/dsinte^S}
\end{equation}
Here, the right hand side is not evaluated for the classical solution,
but it is done so for the broken-zero-mode configurations 
\eqref{BZ} and \eqref{AdS-SOL-3}.
This fact is indicated in \eqref{<WO>=d/dsinte^S} as ``/\!\!\!0-modes''.

Since the expressions \eqref{dh/ds-2} and \eqref{dh/ds-3} 
for the functional derivatives are still valid,
we just substitute the configuration \eqref{AdS-SOL-3} and \eqref{BZ} 
into them and obtain the following expression for 
the numerator of \eqref{<WO>=d/dsinte^S}:\footnote{
The derivative of the boundary term again vanishes.
See appendix \ref{boundary terms}.
}
\begin{align}
& \int{d \Omega_3 \over 2 \pi^2}
{\rm e}^{-S[0]|_{\text{/\!\!\!0-modes}}}
\bigg[
-{\sqrt \lambda \over 4 \pi}
\int d \tau d\sigma
\bigg(
{\delta \over \delta s_0^J(\vec x_J)} 
h_{MN} 
\bigg)
\partial_a {\cal X}^M
\partial^a {\cal X}^N
\bigg]
\Bigg|_{\text{/\!\!\!0-modes}} \notag \\
& \quad =
\int {d \Omega_3 \over 2 \pi^2}
{\rm e}^{\cos \alpha \sqrt{\lambda'}}
\bigg[
{\sqrt \lambda \over 2}
c
{4J \over \ell^{2J}} a^J 
\int d \sigma
\bigg(
{1 \over \cosh^2 \sigma} - f^2(\sigma,\alpha)
\bigg) \tanh^J \sigma {\cal Y}_J 
\bigg]\,.
\label{1/c^2-f^2}
\end{align}
The spherical harmonics ${\cal Y}_J$ is also evaluated 
by using \eqref{BZ} as 
\begin{align}
 {\cal Y}_J &= {f^J(\sigma,\alpha) \over 2^{J/2}}
\Big(
\cosh \sigma_0 \sinh \sigma \cos \alpha
+
\sinh \sigma_0 \cosh \sigma
+
i \sinh \sigma \sin \alpha \cos \beta
\Big)^J \\
 & \sim { 1 \over 2^{J/2}} \tanh^J \sigma 
 ( \cos \alpha + i \sin \alpha \cos \beta)^J
 +
 {\cal O}(\tanh \sigma_0)\,.
\end{align}
Then the $\sigma$-integral reduces to the one in \eqref{S-Y} and 
we are left with the following integral over the broken zero modes:
\begin{equation}
 {1 \over 2 N} {a^J \over \ell^{2J}} \sqrt{J\lambda'} 
 \int {d \Omega_3 \over 2 \pi^2}
{\rm e}^{\cos \alpha \sqrt{\lambda'}}
\cos \alpha
\Big(
 \cos \alpha + i \sin \alpha \cos \beta
\Big)^J.
\end{equation}
The integral with respect to $\beta$ and $\gamma$
results in the following $\alpha$ integral:
\begin{align}
 {1 \over 2N} {a^J \over \ell^{2J}} \sqrt{J \lambda'}
 \times
 {2 \over \pi} {1 \over J+1} \int_0^\pi d \alpha
 {\rm e}^{\cos \alpha \sqrt{\lambda'}}
 \sin \alpha \cos \alpha \sin ((J+1) \alpha)\,. 
\label{int d alpha}
\end{align}
By integrating with respect to $\alpha$\,,
we obtain
\begin{align}
 \eqref{int d alpha} = 
 {1 \over N} {a^J \over \ell^{2J}} \sqrt J
 I_J(\sqrt{\lambda'})
 \bigg(
 1
 - 
 {J+2 \over \sqrt{\lambda'}} 
 { I_{J+1}(\sqrt{\lambda'}) \over I_J(\sqrt{\lambda'}) }
 \bigg)\,.
 \label{semifinal result}
\end{align}
Here, the following equation is used:
\begin{equation}
{1 \over \pi (J+1)}
\int_0^\pi d \alpha
{\rm e}^{\cos \alpha z}
\sin \alpha \cos \alpha
\sin ((J+1)\alpha)
=
{1 \over z } I_J( z )
-
{J+2 \over z^2 }
I_{J+1} (z)\,,
\label{Integral formula}
\end{equation}
which is proved in appendix \ref{Bessel-Functions}.
Finally, dividing \eqref{semifinal result} by 
\eqref{Bessel} we obtain the following result:
\begin{equation}
\eqref{<WO>=d/dsinte^S} =
{1 \over 2N} 
{a^J \over \ell^{2J}}
\sqrt{J \lambda'}
{I_J (\sqrt{\lambda'}) \over I_1 (\sqrt{\lambda'})}
\bigg(
1 - {J+2 \over \sqrt{\lambda'}} 
{I_{J+1}(\sqrt{\lambda'}) \over I_J(\sqrt{\lambda'})}
\bigg)\,.
\label{final result}
\end{equation}

\eqref{final result} is the main result in this paper. 
Since $J$ is an integer satisfying $J \geq 2$\,,
and also since the modified Bessel functions satisfy the relation 
$I_J(\sqrt{\lambda'}) \geq I_{J+1}(\sqrt{\lambda'})$
for these values of $J$\,, the second term in the round bracket 
is subleading in the limit $J / \sqrt{\lambda'} \ll 1$\,.
This means that \eqref{final result} 
reproduces the gauge theory result in that limit.
In fact, deviations in other range is expected,
because the string world sheet we consider does not 
carry any angular momentum and the conservation of it is broken. 
Only in the limit $J \ll \sqrt{\lambda'}$\,,
the world sheet configuration is approximately acceptable.
Hence it is natural that our computation in the gravity side
reproduces the gauge theory result only in the limit.
In order to go beyond such an approximation,
we need to consider world sheet configurations
by taking account of effects of the operator insertion.
Such an analysis for the classical solution is 
given in \cite{Zarembo:2002ph}\cite{Giombi:2012ep}\cite{Enari:2012pq}.
It would be interesting future work to study such effects 
on the broken-zero-mode configurations.

We would like to emphasize that 
\eqref{final result} {\it does not}
give the exact form of the expected modified Bessel function.
When we take the limit $J/\sqrt{\lambda'} \ll 1$ 
in \eqref{final result}, it affects not only the second term 
in the round bracket but also the ratio 
of the modified Bessel function 
$I_J(\sqrt{\lambda'})/I_1(\sqrt{\lambda'})$ itself.
Then the result is not equal to the gauge theory result 
\eqref{<WO>/<W>}, but the limit of it.

Fortunately, the limit $J / \sqrt{\lambda'} \ll 1$\,
still allows the range beyond the purely classical limit. 
An example is the limit in which $\sqrt{\lambda'}$ is 
taken to be large while the combination 
$J^2 / \sqrt{\lambda'} $ is kept finite.
This limit is suggested from the following asymptotic expansion 
of the modified Bessel function \cite{FormulaBook-1}\cite{FormulaBook-2}:
\begin{equation}
I_J(z)
\sim 
{ {\rm e}^z \over \sqrt{ 2 \pi z } }
\sum_{n=0}^\infty
(-1)^n 
{
\Gamma(J+n+{1 \over 2} ) 
\over 
(2z)^n n! \Gamma(J - n + {1 \over 2})
}
+
{{\rm e}^{-z \pm (J + {1 \over 2})\pi i} \over \sqrt{2 \pi z} }
\sum_{n=0}^\infty
{
\Gamma(J+n+{1 \over 2} ) 
\over 
(2z)^n n! \Gamma(J - n + {1 \over 2})
}\,.
\label{asymp-IJ}
\end{equation}
The ratio of the Gamma functions is given by
\begin{equation}
{
\Gamma(J + n + {1 \over 2}) 
\over 
(2z)^n
n! \Gamma(J - n +{1 \over 2} ) 
}
=
{1 \over 2^n n!}
{
J^2 - {1^2 \over 2^2}
\over 
z}
{
J^2 - {3^2 \over 2^2}
\over 
z
}
\cdots
{
J^2 - {(2n-1)^2 \over 2^2}
\over 
z
}
\sim
{1 \over n!} 
\Big( {J^2 \over 2z} \Big)^n\,.
\label{Gamma/Gamma}
\end{equation}
Here, in the last expression, 
we took the large $z$ limit
with keeping the combination $J^2/z$ finite.
If we use \eqref{Gamma/Gamma} in the summation of
\eqref{asymp-IJ}, we obtain
\begin{equation}
I_J(z) \sim 
{{\rm e}^z \over \sqrt{2 \pi z}}
\sum_{n=0}^\infty 
{1 \over n!} 
\bigg(
- {J^2 \over 2z}
\bigg)^n
+
{{\rm e}^{-z \pm (J+{1 \over 2})\pi i} \over \sqrt{2 \pi z}}
\sum_{n=0}^\infty 
{1 \over n!} 
\bigg(
{J^2 \over 2z}
\bigg)^n
=
{ {\rm e}^{z - {J^2 \over 2z}} \over \sqrt{2 \pi z} }
+
{ {\rm e}^{-z \pm (J+{1 \over 2})\pi i +{J^2 \over 2z}} \over \sqrt{2 \pi z}}
\,.
\end{equation}
Then \eqref{final result} reduces 
to the following expression in the limit:
\begin{equation}
 \eqref{final result} \sim 
 {1 \over 2N} {a^J \over \ell^{2J}} \sqrt{J \lambda'}
 {\rm e}^{ - {J^2 \over 2 \sqrt{\lambda'}}}\,.
 \label{asymp}
\end{equation}
Here, we have kept only the leading term 
in the asymptotic expansion of the modified Bessel functions.

The same limit of the modified Bessel function
is considered previously in the case of 
the 1/2 BPS Wilson loop in \cite{Zarembo:2002ph}.
In that paper, the string world sheet action is expanded 
with respect to small fluctuations to quadratic order. 
By integrating the fluctuations, 
the result corresponding to \eqref{asymp}, 
with a replacement $\lambda' \to \lambda$\,,
is reproduced.
It would be interesting to consider the small fluctuations
in the case of the 1/4 BPS Wilson loop and study the relation 
between these two approaches.

\section{Summary and discussions}
\label{Summary and discussions}
In this paper we studied broken zero modes of string world sheet
which exist in the case of the gravity dual 
of the 1/4 BPS Wilson loop operator. 
As proposed in \cite{Drukker:2006ga}, 
we consider the limit $\cos \theta_0 \to 0$, $\lambda \to \infty$ 
with keeping $\lambda' = \lambda \cos^2 \theta_0$ finite.
In this limit, the broken zero modes give 
significant contributions to the string path integral, 
while the effects of other generic fluctuations are 
expected to be negligible.

We started by giving an explicit form of the configuration
of the broken zero modes which depends 
on three parameters $\alpha$, $\beta$ and $\gamma$\,.
The configuration satisfies the correct boundary conditions and 
the Virasoro constraints.
It also reduces to the known solutions 
at $\alpha =0$\,, $\pi$\,.
In the case with $\cos \theta_0 = 0$\,, 
the three parameters are identical to the exact zero modes.
Since we take the small $\cos \theta_0$ limit, 
the leading corrections for the string configuration
with respect to $\cos \theta_0$\,
would be enough to compute the string path integral.
However, having an explicit smooth configuration,
which includes all order terms of the small parameter $\cos \theta_0$, 
makes our arguments more convincing.
Another property of the configuration is that the five parameters 
$(\tau, \sigma, \alpha, \beta, \gamma)$ form a smooth 
coordinate system of the S$^5$.
This clearly shows that two configurations with different 
values of parameters  $(\alpha\,,\,\beta\,,\,\gamma)$ describe
different configurations of the world sheet, i.e., they are not related 
by any redefinition of the world sheet coordinates ($\tau$, $\sigma$)\,.
We have checked that our configuration reproduces 
the modified Bessel function as found in \cite{Drukker:2006ga}.

Since our argument for the explicit configuration is completely ad hoc,
we also tried to derive the form of the broken zero modes 
more systematically by imposing appropriate conditions on the world sheet.
For simplicity, we assumed an ansatz that respects 
the S$^1$ symmetry corresponding to $\tau$\,, 
and also the S$^2$ flat directions of $\beta$ and $\gamma$\,.
This allows three arbitrary functions of $\sigma$ and $\alpha$\,.
Although the conditions on the world sheet 
still leave one arbitrary function undetermined, 
we find that the result is not affected by the choice of it.

In section 4, we studied the gravity dual of 
the correlation function between the 1/4 BPS Wilson loop operator 
and the 1/2 BPS local operator by taking account of the broken 
zero modes. 
The resulting expression \eqref{final result}
agrees with the gauge theory result only in the limit
$J \ll \sqrt{\lambda'}$\,.
Deviation at the outside of the range is attributed to the fact
that the world sheet we consider does not carry any
angular momentum. 
In other words, the string configuration is determined
by neglecting the effects of the operator insertions.
Hence the configuration is valid only in the limit
$J \ll \sqrt{\lambda'}$\,, in which the effects of the
angular momentum is negligible.
Fortunately, the limit $J \ll \sqrt{\lambda'}$ still 
allows a check of the duality beyond the purely classical analysis.
By considering the limit $J/\sqrt{\lambda'} \ll 1$
with keeping the combination $J^2 / \sqrt{\lambda'}$ finite,
we obtain the result \eqref{asymp} which includes 
non-trivial effects of the broken zero modes.

There are several points left for future works.
One is how to study the case with finite $\sqrt{\lambda'}$\,.
For this purpose, we need to consider 
the string world sheet which carries 
the angular momentum. 
The string solutions used in 
\cite{Zarembo:2002ph}\cite{Giombi:2012ep}\cite{Enari:2012pq}
could be a good starting point for such studies.
As mentioned at the end of the previous section, 
studying the relation between the method of the broken zero modes 
and the one in \cite{Zarembo:2002ph} would be also 
interesting future work.
Another point which is not quite clear to the author 
is the treatment of the path integral measure and the
derivative with respect to the source $s_0^J$\,.
Since the S$^3$ measure for the broken zero modes
comes from the target space geometry, it would
depend on the source.
In that case it seems to be not quite clear
whether the derivative should be from the outside of the 
path integral and it hits the measure, or it is just inside
the path integral. 
In the present paper, we took the latter choice as an assumption.
We would like to address this issue in the future work.

\section*{Acknowledgment}
The author of the present paper would like to 
thank 
Y.~Imamura, 
K.~Ito, 
T.~Kimura,
K.~Sakai, 
Y.~Satoh, 
S.~Yamaguchi, 
K.~Yoshida
and also all members of the particle theory group in CST, Nihon University, 
for discussions, comments and encouragements.
He also would like to thank the Yukawa Institute 
for Theoretical Physics at Kyoto University. 
Discussions during the workshop YITP-W-14-4
``Strings and Fields'' were 
useful to complete this work.

\appendix

\section{Boundary terms}
\label{boundary terms}
In the present case, 
the boundary term proposed in \cite{Drukker:1999zq} 
can be written as follows:
\begin{equation}
S_{\rm boundary} = 
-
\int_0^{2 \pi} d \tau {\partial {\cal L} \over \partial (\partial_\sigma z)} z
\Bigg|_{\sigma = \sigma_{\rm min}}
=
- 
{\sqrt \lambda \over 2 \pi}
\int_0^{2 \pi} d \tau 
\big( g_{z \mu}^{\rm AdS} + h_{z \mu}^{\rm AdS} \big) \partial_\sigma {\cal X}^\mu z \Bigg|_{\sigma=\sigma_{\rm min}}\,.
\end{equation}
Here, $g^{\rm AdS}_{\mu \nu}$ is the AdS$_5$ metric and 
$h^{\rm AdS}_{\mu \nu}$ is the fluctuation which depends on the source 
$s_0^J$ of the local operator.
If we set $s_0^J = 0$\,, then the fluctuation $h^{\rm AdS}_{\mu \nu}$ 
is zero, and by using \eqref{AdS-SOL-3}, 
the boundary term is evaluated as
\begin{equation}
S_{\rm boundary} \big|_{s_0^J=0} = 
-
{\sqrt \lambda \over 2 \pi}
\int_0^{2\pi} d \tau
{1 \over z} \partial_\sigma z \Bigg|_{\sigma = \sigma_{\rm min}}
=
-{\sqrt \lambda \over \sinh \sigma_{\rm min} \cosh \sigma_{\rm min}}\,.
\end{equation}
This term exactly cancels the first term of \eqref{S(a,b,c)-0} 
in the limit $\sigma_{\rm min} \to 0$.

Next let us check that the functional derivative with respect to
$s_0^J(\vec x_J)$ vanishes.
In \eqref{dh/ds} and \eqref{1/c^2-f^2}, 
their could be an additional term like
\begin{align}
- {\delta \over \delta s_0^J(\vec x_J)} S_{\rm boundary}
& =
{\sqrt \lambda \over 2 \pi}
\int_0^{2 \pi} d \tau 
\bigg(
{\delta \over \delta s_0^J(\vec x_J)} h_{z \mu}^{\rm AdS} 
\bigg)
\partial_\sigma {\cal X}^\mu z \Bigg|_{\sigma = \sigma_{\rm min}} \notag \\
& \sim 
{\sqrt \lambda \over 2 \pi}
\int_0^{2 \pi} d \tau 
\bigg(
2J {1 \over z^2} c {z^J \over \ell^{2J}} {\cal Y}_J 
\bigg)
\partial_\sigma z z
\Bigg|_{\sigma = \sigma_{\rm min}} \notag \\
& = 
{1 \over 2N}
\sqrt{J \lambda} (J+1)
{a^J \over \ell^{2J}}
(\cos \alpha + i \sin \alpha \cos \beta)^J
{\tanh^{2J-1} \sigma_{\rm min} \over \cosh^2 \sigma_{\rm min}}\,.
\end{align}
From the final expression, we see that
it vanishes in the limit $\sigma_{\rm min} \to 0$\,,
since $J$ is an integer greater than or equal to $2$\,.

\section{Properties of  the modified Bessel functions}
\label{Bessel-Functions}
Let us explain some properties of the modified Bessel function
$I_J(z)$ which we use in the main text. 
In the following discussion, 
we assume that $J$ is a non-negative integer
and $z \geq 0$\,.

The first point we explain 
is the $\alpha$-integral in \eqref{Integral formula},
which  may not be very clear.
We start with the following integral formula 
for the modified Bessel function:
\begin{equation}
 I_J(z) = 
 {1 \over \pi}
 \int_0^\pi d \alpha \cos (J \alpha) {\rm e}^{\cos \alpha z}\,,
 \quad (J=0,1,\ldots)\,.
\label{I_J=int}
\end{equation}
We can confirm this expression by checking that it satisfies
the following recurrence formulas of the modified Bessel functions
\begin{align}
\begin{split}
 &I_J(z) - I_{J+2}(z) = {2 (J+1) \over z} I_{J+1}(z)\,,\\ 
 & I_J(z) + I_{J+2}(z) = 2 I_{J+1}'(z)\,, 
\end{split}
\label{I_J pm I_J}
\end{align}
and also it satisfies the initial condition 
\begin{equation}
I_0(z) = {1 \over \pi} \int_0^\pi d \alpha {\rm e}^{\cos \alpha z}\,.
\end{equation}
Next by integrating by part and 
using \eqref{I_J=int}, we find the following relation:
\begin{align}
 & {1 \over \pi(J+1) }
 \int_0^\pi d \alpha {\rm e}^{\cos \alpha z}
 \sin \alpha \sin ( (J+1) \alpha ) \notag \\
 & \qquad =
  - {1 \over \pi(J+1) z}
 \int_0^\pi d \alpha 
 {\partial \over \partial \alpha} 
 \Big( {\rm e}^{\cos \alpha z} \Big)
 \sin ( (J+1) \alpha ) \notag \\
&\qquad  =
 {1 \over z} I_{J+1}(z)\,.
 \label{J/z I_J}
\end{align}
Finally we differentiate the first and the last expression
of \eqref{J/z I_J} with respect to $z$ 
and obtain \eqref{Integral formula} as 
\begin{align}
& {1 \over \pi (J+1)}
\int_0^\pi d \alpha 
{\rm e}^{\cos \alpha z}
\sin \alpha \cos \alpha
\sin ((J+1) \alpha) 
\notag \\ & \qquad 
=
- { 1 \over z^2} I_{J+1}(z)
+ { 1 \over z} I_{J+1}'(z) \notag \\
& \qquad =
{1 \over z}I_J(z)
-
{J+2 \over z^2} I_{J+1}(z)\,.
\end{align}
Here, the above recurrence formulas \eqref{I_J pm I_J} 
are again used when we go to the final expression.

The recurrence formulas can be also used to show 
the following inequality for $z \geq 0$\,:
\begin{equation}
I_J(z) \geq I_{J+1} (z) \quad (J=0,1,\ldots )\,.
\label{IJ>IJ+1}
\end{equation}
First, from \eqref{I_J=int} we have
\begin{equation}
I_0(z) - I_1(z) = 
{1 \over \pi}
\int_0^\pi d \alpha 
( 1 - \cos \alpha )
{\rm e}^{\cos \alpha z}
>
0\,.
\label{I0-I1}
\end{equation}
Next, from the second equation of \eqref{I_J pm I_J}, we find
\begin{equation}
{d \over d z} \big( I_{J+1}(z)  - I_{J+2}(z) \big)
=
{1 \over 2}
\bigg[
\big( I_J (z) - I_{J+1} (z) \big)
+
\big( I_{J+2} (z)- I_{J+3} (z) \big)
\bigg]\,.
\label{d/dz(IJ-IJ+1)}
\end{equation}
From \eqref{I0-I1}, \eqref{d/dz(IJ-IJ+1)} 
and the initial conditions\footnote{
For the range $0< z < 2$, we find
$ I_J(z) - I_{J+1}(z) > 
  I_J(z) - {2(J+1) \over z} I_{J+1}(z)
  = I_{J+2}(z) > 0 \,\, (J = 0,1,\ldots)$\,,
which can be used as alternative
initial conditions.
}
$I_0(0)=1$ and $I_J(0) = 0$ ($J=1,2,\ldots$)\,,
we see that the relation \eqref{IJ>IJ+1} holds 
for $z \geq 0$.

\section{Supersymmetry}
\label{Supersymmetry}
Supersymmetries preserved by the system including 
a 1/4 BPS Wilson loop and a 1/2 BPS local operator
is studied in \cite{Semenoff:2006am}.
Here, we review the analysis with taking account of 
the distance $\ell$ between the local operator and 
the center of the Wilson loop operator.

The supersymmetries preserved by the Wilson loop operator 
\eqref{W(C)}, with \eqref{x} and \eqref{Theta},
are given by the following conditions \cite{Drukker:2006ga}:
\begin{equation}
\Big[
- i \sin \tau \gamma_1 
+ i \cos \tau \gamma_2
+ \sin \theta_0 \cos \tau \gamma_5 
+ \sin \theta_0 \sin \tau \gamma_6
+ \cos \theta_0 \gamma_7
\Big]
\Big[
\epsilon_0 + 
a 
\big(
\cos \tau \gamma_1 
+
\sin \tau \gamma_2
\big)
\epsilon_1
\Big]=0\,.
\label{W-SUSY}
\end{equation}
Here, we use the ten-dimensional notation.
$\gamma_M$ ($M=1, \ldots , 10$) 
are the ten-dimensional gamma matrices
which satisfy $\{ \gamma_M , \gamma_N \} = 2 \delta_{MN}$.
$M=1,2,3,4$ correspond to the four-dimensional space 
of the SYM theory, while $M=5,\ldots,10$ correspond to 
the reduced dimensions.
Each spinor $\epsilon_0$ and $\epsilon_1$ has 
an opposite chirality, in the sense of ten-dimension,
and generates the Poincar\'e supersymmetry 
and the conformal supersymmetry, respectively.
These two spinors appear in the transformation of the 
bosonic field only in the following combination:
\begin{equation}
\epsilon = \epsilon_0 + x^i \gamma_i \epsilon_1\,.
\end{equation}

The independent conditions coming from \eqref{W-SUSY}
are summarized in \cite{Drukker:2006ga}.
For $\theta_0 = 0$, 
\eqref{W-SUSY} relates $\epsilon_0$ and $\epsilon_1$ 
as follows:
\begin{equation}
\epsilon_0 = i a \gamma_1 \gamma_2 \gamma_7 \epsilon_1\,.
\label{W-SUSY-0}
\end{equation}
On the other hand, for the other special case with $\theta_0 = \pi/2$\,, 
each spinor $\epsilon_0$ and $\epsilon_1$ satisfies the following 
common conditions independently:
\begin{equation}
\big(1 - i \gamma_2 \gamma_5 \big) \epsilon_0
=
\big(1  +i \gamma_1 \gamma_6 \big) \epsilon_0
=
0\,,\quad 
\big(1 - i \gamma_2 \gamma_5 \big) \epsilon_1
=
\big(1 + i \gamma_1 \gamma_6 \big) \epsilon_1
=
0\,. 
\label{W-SUSY-Pi/2}
\end{equation}
Finally, for a generic value of $\theta_0$\,, 
\eqref{W-SUSY} reduces to the following conditions:
\begin{equation}
\cos \theta_0 \epsilon_0
=
a
\big(
- i \gamma_1 + \sin \theta_0 \gamma_6 
\big)
\gamma_7 \gamma_2 
\epsilon_1\,, \qquad
\big(
1 - \gamma_1 \gamma_2 \gamma_5 \gamma_6
\big)
\epsilon_1
=0\,.
\label{W-SUSY-generic}
\end{equation}
In addition to these ``Wilson loop conditions'',
we need to impose the following conditions
coming from the local operator ${\cal O}_J$
located at $\vec x = (0,0,0,\ell)$:
\begin{equation}
\Big[
1 + i \gamma_7 \gamma_8
\Big]
\Big[
\epsilon_0
+
\ell \gamma_4 \epsilon_1
\Big]
=0\,.
\label{O_J-SUSY}
\end{equation}
Let us summarize the combined conditions for each 
value of $\theta_0$\,.
\begin{itemize}
\item $\theta_0 = 0$\,: \\ 
From \eqref{W-SUSY-0} and \eqref{O_J-SUSY}, 
we find that the following condition is 
imposed on $\epsilon_1$ 
\begin{align}
\Big[
1 + i \gamma_7 \gamma_8
\Big]
\Big[
i a \gamma_1 \gamma_2 \gamma_7 + \ell \gamma_4
\Big] 
\epsilon_1
=
0 \,.
\label{P_1}
\end{align}
Since the matrix
$(ia\gamma_1 \gamma_2 \gamma_7 + \ell \gamma_4)$
is invertible, the condition \eqref{P_1} projects out 
half of the degrees of freedom (d.o.f.) of $\epsilon_1$\,,
and then 8 of them survive.
Since $\epsilon_0$ is determined by \eqref{W-SUSY-0},
the whole system including $W(C)$ and ${\cal O}_J$
preserves $8$ supersymmetries.

\item $\theta_0 = \pi/2$

In this case, we may recombine 
$\epsilon_0$ and $\epsilon_1$ as
\begin{equation}
\begin{cases}
\epsilon_+ = \epsilon_0 + \ell \gamma_4 \epsilon_1 \,, \\[4mm]
\epsilon_- = \epsilon_1 - \ell \gamma_4 \epsilon_0 \,,
\end{cases}
\quad \leftrightarrow \quad 
\begin{cases}
\epsilon_0 = 
\displaystyle{1 \over 1 + \ell^2} 
\big( \epsilon_+ - \ell \gamma_4 \epsilon_- \big)\,, \\[4mm]
\epsilon_1 = 
\displaystyle{1 \over 1 + \ell^2}
\big(\epsilon_- + \ell \gamma_4 \epsilon_+ \big)\,,
\end{cases}
\end{equation}
where the each spinor $\epsilon_\pm $ has 16 d.o.f.
with an opposite chirality.
Then the condition \eqref{W-SUSY-Pi/2} can be
equivalently imposed on $\epsilon_\pm$ as
\begin{equation}
(1 - i \gamma_2 \gamma_5) \epsilon_\pm
=
(1 + i \gamma_1 \gamma_6) \epsilon_\pm
=
0\,.
\end{equation}
These conditions allow only 
quarter of the each spinor $\epsilon_\pm$\,
and $4 + 4 = 8$ d.o.f. survive.
Then the additional condition \eqref{O_J-SUSY} 
reduces the d.o.f. of $\epsilon_+$
to the half of it, while $\epsilon_-$ is not affected.
Then the whole system preserves 
$4  + {1 \over 2} \times 4 = 6$ supersymmetries.

\item $ 0 < \theta_0 < \pi/2$ \\
From the first condition of \eqref{W-SUSY-generic} and \eqref{O_J-SUSY}, 
$\epsilon_1$ needs to satisfy the following condition:
\begin{equation}
\Big(
1 + i \gamma_7 \gamma_8
\Big)
\Gamma
\epsilon_1 
=
0\,,
\quad
\big(\,
\Gamma \equiv ai\gamma_1 \gamma_2 \gamma_7 
+ a \sin \theta_0 \gamma_2 \gamma_6 \gamma_7 
+ \ell \cos \theta_0 \gamma_4\,
\big)
\,.
\label{P_2}
\end{equation}
Since the matrix $\Gamma$ is invertible, 
\eqref{P_2} projects out half of the d.o.f.
of $\epsilon_1$ and 8 of them survive.
This is a generalization of the case with 
$ \theta_0 = 0 $\,.
Now in the case of generic $\theta_0$\,, $\epsilon_1$ 
needs to satisfy the second condition of \eqref{W-SUSY-generic}\,.
Since $\Gamma $ and $\gamma_1 \gamma_2 \gamma_5 \gamma_6$ commute, 
we can impose the following equivalent condition:
\begin{equation}
(1-\gamma_1 \gamma_2 \gamma_5 \gamma_6) \Gamma \epsilon_1=0\,.
\label{P_3}
\end{equation}
Then, since $i\gamma_7 \gamma_8$ 
and $\gamma_1 \gamma_2 \gamma_5 \gamma_6$ commute, 
two conditions \eqref{P_2} and \eqref{P_3} 
allow ${1 \over 2}\times {1 \over 2} \times 16 = 4$ 
independent d.o.f. of the original spinor $\epsilon_1$\,.
The other spinor $\epsilon_0$ is determined by 
the first equation of \eqref{W-SUSY-generic} and the whole system
preserves $4 = {1 \over 8} \times (16+16)$ supersymmetries.

\end{itemize}

\end{document}